\newif\ifLNCS
\LNCStrue

\ifLNCS
    \documentclass{llncs}
    \usepackage{amsmath,amssymb}
\else
    \documentclass[a4paper,11pt]{article}
    \usepackage{amsmath,amsthm,amssymb}
    \usepackage[top=1in, bottom=1in, left=1in, right=1in]{geometry}
\fi

\usepackage[T1]{fontenc}
\usepackage{xspace}
\usepackage[dvipsnames]{xcolor}
\usepackage{hyperref}
\usepackage{mathtools}
\usepackage{enumitem}

\setlist{noitemsep}
\setlist[2]{noitemsep}

\ifLNCS
\else
\newtheorem{defin}{Definition}
  
\newtheorem{theo}[defin]{Theorem}
  \newenvironment{theorem}{\begin{theo} \sl}{\end{theo}}
\newtheorem{lem}[defin]{Lemma}
  \newenvironment{lemma}{\begin{lem} \sl}{\end{lem}}
  
\newtheorem{propo}[defin]{Proposition}
  \newenvironment{proposition}{\begin{propo} \sl}{\end{propo}}
\newtheorem{coro}[defin]{Corollary}
  
\newtheorem{obse}[defin]{Observation}
  
\newtheorem{conj}[defin]{Conjecture}
  \newenvironment{conjecture}{\begin{conj} \sl}{\end{conj}}
\newtheorem{rem}[defin]{Remark}
  
\newtheorem{myfact}[defin]{Fact}
  
\fi

\newenvironment{myproof}{\emph{Proof.}}{\hfill $\Box$ \medskip\\}

\definecolor{ablue}{rgb}{0.3,0.4,0.8}
\definecolor{ared}{rgb}{0.95,0.4,0.4}
\definecolor{agreen}{rgb}{0,0.5,0.25}
\definecolor{ayellow}{rgb}{0.95,0.85,0.3}

\newcommand{\Reals}{\mathbb{R}}

\newcommand{\cG}{\mathcal{G}}
\newcommand{\graph}{\cG}

\newcommand{\cC}{\mathcal{C}}
\newcommand{\cP}{\mathcal{P}}

\newcommand{\cL}{\mathcal{L}}

\newcommand{\cX}{\mathcal{X}}

\newcommand{\eps}{\varepsilon}

\newcommand{\eqdef}{:=}

\DeclareMathOperator{\dist}{dist}

\DeclareMathOperator{\radius}{radius}
\newcommand{\myin}{\mathrm{in}}
\newcommand{\myout}{\mathrm{out}}

\newcommand{\mytrue}{\mbox{{\sc True}}}
\newcommand{\myfalse}{\mbox{{\sc False}}}

\DeclareMathOperator{\ball}{ball}
\DeclareMathOperator{\cent}{center}

\renewcommand{\leq}{\leqslant}
\renewcommand{\geq}{\geqslant}

\newcommand{\etal}{\emph{et~al.}\xspace}
\newcommand{\Otilde}{\widetilde{O}}

\newcommand{\nph}{{\sc np}-hard\xspace}
\newcommand{\IS}{\textsc{Independent Set}\xspace}
\newcommand{\DS}{\textsc{Dominating Set}\xspace}
\newcommand{\wDS}{\textsc{Weighted Dominating Set}\xspace}
\newcommand{\CDS}{\textsc{Connected Dominating Set}\xspace}
\newcommand{\CornotDS}{\textsc{(Connected) Dominating Set}\xspace}
\newcommand{\CVC}{\textsc{Connected Vertex Cover}\xspace}
\newcommand{\FVS}{\textsc{Feedback Vertex Set}\xspace}
\newcommand{\CFVS}{\textsc{Connected Feedback Vertex Set}\xspace}
\newcommand{\CornotFVS}{\textsc{(Connected) Feedback Vertex Set}\xspace}
\newcommand{\HC}{\textsc{Hamiltonian Cycle}\xspace}
\newcommand{\ST}{\textsc{Steiner Tree}\xspace}

\newcommand{\threeSAT}{$3$-\textsc{sat}\xspace}
\newcommand{\niceSAT}{$(3,3)$-\textsc{sat}\xspace}

\newcommand{\MIF}{\textsc{Maximum Induced Forest}\xspace}

\newcommand{\BeginMyItemize}{\begin{itemize}\setlength{\itemsep}{-\parskip}}
\newcommand{\EndMyItemize}{\end{itemize}}
\newcommand{\myitemize}[1]{\BeginMyItemize #1 \EndMyItemize}

\ifLNCS
    \title{Lower Bounds for Dominating Set in Ball Graphs and for Weighted Dominating Set in Unit-Ball Graphs\thanks{This research was supported by the Netherlands Organization for Scientific Research NWO under project no. 024.002.003 ({\sc Networks}).}}
    \titlerunning{Lower Bounds for Dominating Set}

    \author{Mark de Berg\inst{1} \and S\'andor Kisfaludi-Bak\inst{2}}
    \institute{TU Eindhoven, The Netherlands, \email{m.t.d.berg@tue.nl}
               \and Max Planck Institut f\"ur Informatik, Saarbr\"ucken, Germany, \email{sandor.kisfaludi-bak@mpi-inf.mpg.de}}
    \authorrunning{M. de Berg and S. Kisfaludi-Bak}
\else
    \title{Lower Bounds for Dominating Set in Ball Graphs and for Weighted Dominating Set in Unit-Ball Graphs}

    \author{Mark de Berg\thanks{Eindhoven University of Technology, The Netherlands; \url{m.t.d.berg@tue.nl}, Supported by the Netherlands Organization for Scientific Research NWO under project no. 024.002.003.} \and S\'andor Kisfaludi-Bak$\thanks{Max Planck Institute for Informatics, Saarbr\"ucken, Germany; \url{sandor.kisfaludi-bak@mpi-inf.mpg.de}}$}
    
    \date{}
\fi

\begin{document}

\maketitle

\begin{abstract}
Recently it was shown that many classic graph problems---\IS, \DS, \HC, and more---can
be solved in subexponential time on unit-ball graphs.
More precisely, these problems can be solved in $2^{O(n^{1-1/d})}$ time on unit-ball graphs
in $\Reals^d$,  which is tight under ETH.
The result can be generalized to
intersection graphs of similarly-sized fat objects.

For \IS the same running time can be achieved for non-similarly-sized fat objects, and for the weighted version of the problem.
We show that such generalizations most likely are not possible for \DS: assuming ETH, we prove that
\myitemize{
\item
there is no algorithm with running time $2^{o(n)}$ for \DS on (non-unit) ball graphs in $\Reals^3$;
\item
there is no algorithm with running time $2^{o(n)}$ for \wDS on unit-ball graphs in $\Reals^3$;
\item
there is no algorithm with running time $2^{o(n)}$ for
\DS, \CDS, or \ST on
intersections graphs of arbitrary convex (but non-constant-complexity) objects in the plane.
}
\end{abstract}

\section{Introduction}
\begin{quotation}
\noindent \emph{Over the past few years the authors had the privilege to collaborate with
Hans Bodlaender on various algorithmic problems on geometric intersection graphs.
In this short paper, written on the occasion of Hans's 60th birthday,
we further explore some of the research directions inspired
by this joint  research. Hans, we hope you will enjoy reading the paper and
are looking forward to more collaborations. And of course: happy birthday!}
\end{quotation}
Many classic optimization problems on graphs cannot be solved in subexponential
time---that is, in time $2^{o(n)}$, where $n$ is the number of vertices---on general
graphs, assuming the Exponential-Time Hypothesis~\cite{ethcite}. For planar graphs,
however, subexponential algorithms are often possible. In particular, on planar
graphs one can often obtain $2^{O(\sqrt{n})}$ running time. This is for example
the case for \IS (and, hence, {\sc Vertex Cover}), for \DS, and for \HC~\cite{fptbook}.
The fact that so many \nph problems can be solved in $2^{O(\sqrt{n})}$ time
on planar graphs has been dubbed the \emph{square-root phenomenon}~\cite{squarerootphen}.
Following recent work, we will explore to what extent this phenomenon
can also be observed in certain types of geometric intersection graphs that
generalize planar graphs.

\begin{figure}[t]
\centering
\includegraphics{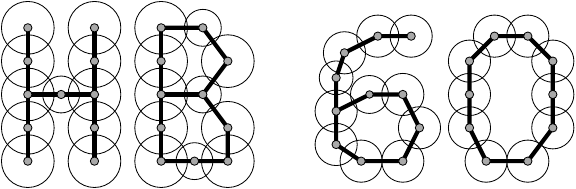}
\caption{An intersection graph of a set of disks in the plane.}\label{fig:hb60}
\end{figure}
The \emph{intersection graph} $\graph[S]$ of a set $S$ of objects in $\Reals^d$ is
the graph whose vertex set corresponds to $S$ and that
has an edge between two vertices if and only if the corresponding two
objects intersect; see Fig.~\ref{fig:hb60}.
It is well known that the class of planar graphs is equivalent to the
class of contact graphs of disks in the plane~\cite{Koebe}, so contact graphs
exhibit the square-root phenomenon. (A \emph{contact graph}
of disks is the intersection graph of a set~$S$ of closed
disks with disjoint interiors.) Does the square-root phenomenon also arise
for other types of intersection graphs? And is there a version
of the square-root phenomenon for intersection graphs in higher dimensions?
The answer is yes: De~Berg~\etal~\cite{bbkmz-fetht-18} (see also Kisfaludi-Bak's thesis~\cite{skb_phd})
recently showed that all the classic
graph problems mentioned above can be solved in $2^{O(\sqrt{n})}$ time
on unit-disk graphs---a \emph{unit-disk graph} is the intersection graph
of unit disks in the plane---, and in $2^{O(n^{1-1/d})}$ time on unit-ball
graphs in $\Reals^d$. More generally, these problems can be solved
in $2^{O(n^{1-1/d})}$ time for intersection graphs of \emph{similarly-sized fat objects}. Here we say an object~$o$ is \emph{fat} if there is a ball~$b_{\myin}\subseteq o$ of
radius $\rho_\myin$ and a ball~$b_{\myout}\supseteq o$
of radius~$\rho_\myout$ such that $\rho_\myin/\rho_\myout\geq \alpha$, where $\alpha$ is an absolute constant, and we say that a collection of objects is \emph{similarly sized}, if the ratio of the largest and smallest object diameter is at most some absolute constant.
\medskip

Algorithms with running time $2^{O(\sqrt{n})}$ on planar graphs are typically
based on the Planar Separator Theorem~\cite{LiptonT79,LiptonT80}. This theorem states that any
planar graph $\graph=(V,E)$ has a balanced separator of size~$O(\sqrt{n})$, that
is, a subset $C\subset V$ of size $O(\sqrt{n})$ whose removal splits $\graph$
into connected components with at most $\delta|V|$ vertices each, for some constant~$\delta<1$. For unit-disk graphs
such a result is clearly impossible, since unit-disk graphs can have arbitrarily
large cliques. However, it is always possible to find a balanced separator that consists of a
small number of cliques. More precisely, De~Berg~\etal proved that any
ball graph (or more generally, intersection graph of fat objects) in $\Reals^d$ admits a balanced separator $C_1 \cup \cdots \cup C_k$
such that $\sum_{i=1}^k \log (|C_i|+1) = O(n^{1-1/d})$ and each $C_i$ is a clique.
This clique-based separator theorem forms the basis of an algorithmic framework~\cite{bbkmz-fetht-18}
for obtaining algorithms with $2^{O(n^{1-1/d})}$ running time for all problems mentioned above.
Interestingly, the framework only
works for \emph{unit-ball} graphs---or, more generally, intersection graphs of
\emph{similarly-sized} fat objects---even though the underlying clique-based separator
exists for arbitrarily-sized balls (or more generally, for fat objects).
An exception is \IS, where the clique-based separator theorem
immediately gives an algorithm with $2^{O(n^{1-1/d})}$ running time for
arbitrarily-sized balls~\cite{bbkmz-fetht-18}. For \IS it is also easy to extend
the result to the weighted version of the problem. In addition, it is possible to relax the fatness assumption while maintaining the subexponential behavior:  for \IS in $\Reals^2$ there is even a subexponential algorithm for arbitrary polygons~\cite{facility}, while for $d\geq 3$ a favorable trade-off can be established between fatness and running time (in case of similarly-sized objects) that is ETH-tight~\cite{fat_impact}. In this paper we explore to what
extent the framework of De~Berg~\etal~\cite{bbkmz-fetht-18} can be generalized. In particular, we study for which problems the restriction
to similarly-sized objects is necessary. We also study whether weighted versions of
the above problems can be solved in subexponential time
for similarly-sized objects.
\medskip

\paragraph{Our results.}
In Section~\ref{sec:nosimsize} we first argue that for \CVC,
\FVS and \CFVS the restriction to similarly-sized objects in not necessary: these problems admit subexponential algorithms for
arbitrarily-sized fat objects in~$\Reals^d$.
After that we show this generalization is not possible
for \DS, \CDS, or \ST: already in the plane there is no
algorithm with running time $2^{o(n)}$ for these problems
on intersection graphs of arbitrarily-sized fat objects, assuming ETH.
The fat objects we use for this lower bound are convex, but
they do not have constant description complexity.

In Sections~\ref{se:LB-DS} and \ref{se:LB-wDS} we then turn to the main topic of this paper: the complexity of
({\sc Weighted}) \DS on ball graphs in $\Reals^3$. (Recall that the \DS problem
is to decide, given a graph $\graph=(V,E)$ and a number $k$, if there is a vertex subset $D\subset V$ of size at most $k$ such that all vertices $v\in V\setminus D$ are adjacent to some vertex in $D$.)
As our main contribution, we show that \DS on arbitrary ball graphs in $\Reals^3$ cannot be solved in $2^{o(n)}$ time, assuming ETH. In addition, we consider \wDS, where each vertex is assigned a real weight, and the goal is to find the dominating set of weight at most~$k$. It turns out that this is considerably harder than the unweighted problem:
even on unit-ball graphs in $\Reals^3$, \wDS cannot be solved in $2^{o(n)}$ time, assuming ETH.



\paragraph{Remark.}
It is known that recognizing if a given graph~$\graph$
is a unit-disk graph is \nph~\cite{bk-udgr-98}, so conceivably certain problems
are easier when the input is a set of balls inducing a ball graph, rather than
just the graph itself.
It is thus desirable to develop algorithms that do not
need to know the set $S$ of objects defining the intersection graph, but
that work with only the graph~$\graph[S]$ as input. De Berg~\etal~\cite{bbkmz-fetht-18}
do this by introducing a variant of treewidth. The concept later dubbed as $\cP$-flattened
treewidth~\cite{skb_phd} is based on the idea of weighted treewidth~\cite{weightedtw}.
The need for finding separators geometrically is then alleviated by using the
treewidth-approximation algorithm of Bodlaender~\etal~\cite{tw-approx}.
On the other hand, lower bounds become stronger when they still apply in the geometric
setting where the set $S$ defining the intersection graph is given. All our
lower bounds have this property. From now on, whenever we speak of a problem on
intersection graphs, we mean this geometric version of the problem. For example,
in the \DS problem on ball graphs, we have as input a set of balls with rational
coordinates and radii, and an integer~$k$, and the question is if the intersection
graph~$\graph[S]$ has a dominating set of size~$k$.

\section{Non-similarly sized fat objects}\label{sec:nosimsize}
In this section we consider intersection graphs of arbitrary fat objects---they are
not restricted to be of similar size, and they can have complicated shapes (in particular,
they need not be balls). We first argue that \CVC, \FVS, and \CFVS have algorithms
with running time~$2^{O(n^{1-1/d})}$ for arbitrary fat objects in~$\Reals^d$, even for the weighted
version of these problems. Then we show that \DS, \CDS, and \ST are more
difficult: there are no subexponential algorithms for
these problems in intersection graphs of fat objects in $\Reals^2$.

\paragraph*{\CVC and \CornotFVS.}
The key component in using the algorithmic machinery of
De~Berg~\etal~\cite{bbkmz-fetht-18} for a given graph problem is to show
that any clique contains at most constantly many vertices from an optimal solution.
For example, a clique contains at most one vertex from a solution to \IS.
For \FVS this is not true. Here, however, we can look at the complementary
problem: instead of finding a minimum-size subset of vertices whose removal destroys
all cycles, we can find a maximum-size subset that forms an induced forest.
Indeed, there is a feedback vertex set of size at most $k$ if and only if there is an induced
forest of size at least $n-k$, so we can concentrate on solving \MIF instead.

For \MIF we have the desired property that any solution contains at most
a constant number of vertices---at most two, to be precise---from a clique.
In order to solve \MIF we can now develop a standard
divide-and-conquer algorithm, using the clique-based separator of~\cite{bbkmz-fetht-18}.
Of course the two sub-problems we get, one for the subgraph
``inside'' the separator and one for the subgraph ``outside'' the separator,
are not independent: a solution on the inside, together with a solution
in the separator itself, puts some constraints on the solution on the
outside. For this reason, one needs to keep track of the connected components
created on each side. The naive algorithm would give a running time of
$2^{O(n^{1-1/d}\log n)}$, which is subexponential, but not optimal. The rank-based
approach of Bodlaender~\etal~\cite{single-exponential} can be applied to
improve this to $2^{O(n^{1-1/d})}$.

For \CVC and \CFVS, we also have the property that at most constantly
many vertices from any clique are present in the complement of a solution:
the complement of a solution to \CVC contains at most one vertex per clique
(since the complement is an independent set), and the complement of a solution
to \CFVS contains at most two vertices per clique (since, as above,
the complement is an induced forest). Using the framework of
De~Berg~\etal~\cite{bbkmz-fetht-18}, and the rank-based approach~\cite{single-exponential}
to handle the connectivity issues, we can again obtain algorithms
with $2^{O(n^{1-1/d})}$ running time.

\paragraph*{\CornotDS and \ST.}
For \CornotDS and \ST one cannot guarantee that the solution
(or its complement) uses only a constant number of vertices from
a clique. Hence, the framework from~\cite{bbkmz-fetht-18} cannot easily be used. As the next theorem states, a subexponential algorithm is unlikely to exist for these problems on
intersection graphs of fat objects. The objects in the proof
will be convex, their fatness will be arbitrarily close to~1, but they will have high (non-constant) complexity.
\begin{theorem}
Let $\eps>0$ be any fixed constant. There is no $2^{o(n)}$ algorithm for \DS, \CDS, or \ST
in intersection graphs of convex $(1-\eps)$-fat objects in $\Reals^2$,
unless the Exponential-Time Hypothesis fails.
\end{theorem}
\begin{myproof}
Our proof works in two steps. We first show that \DS, \CDS, and \ST do
not admit algorithms with running time $2^{o(n)}$ on split graphs (assuming ETH).
(A \emph{split graph} is a graph $\graph=(A\cup B,E)$ on $2n$ vertices
such that $A$ induces a clique and $B$ induces an independent set.)
We then show that any split graph can be realized as the intersection
graph on convex $(1-\eps)$-fat objects.
\medskip

It is known that \DS on general graphs with $n$ vertices does not
admit an algorithm with running time $2^{o(n)}$, assuming ETH~\cite{fptbook}.
This lower bound carries over to split graphs. To see this, consider
an arbitrary graph $\graph=(V,E)$ on $n$ vertices. We create a split graph
$\graph'=(A\cup B,E')$ with $2n$ vertices, where for each $v_i\in V$
we create a vertex $a_i\in A$ and $b_i\in B$. We then add
the edge $(a_i,b_j)$ to $E'$ if and only if $i=j$ or $(v_i,v_j)\in E$.
Finally, we add all edges between vertices in~$A$, so $A$ induces a clique.

Note that $\graph$ has a dominating set of size $k$ if and only if $\graph'$ has a
dominating set of size $k$. To see this, note that a dominating set $D\subseteq V$
for $\graph$
corresponds to a dominating set $D'\subseteq A$ with the same vertex indices for $\graph'$.
On the other hand, if $D'$ is a dominating set in $\graph'$, then any vertex $b_i\in
D'\cap B$ can be exchanged with $a_i$, since $a_i$ dominates a superset of the
vertices dominated by $b_i$. If both $a_i$ and $b_i$ are present
in $D'$, then $b_i$ can be removed from $D'$. Therefore, we can find a
dominating set $D'\subset A$ of size at most $k$ for $\graph'$,
which corresponds to a dominating set of the same size for $\graph$.

From the above we can conclude that there is no $2^{o(n)}$ algorithm for \DS on split graphs,
assuming ETH. Notice that the same statement holds for \CDS, as the created
dominating set $D'\subset A$ is connected. The statement holds for
\ST as well, as we now argue. Recall that an instance of \ST is a graph~$\graph=(V,E)$, a terminal set~$T$,
and a natural number~$k$, and the question is to decide if there is a set
$W\subseteq V$ of size at most $k$ such that $T\cup W$ induces a connected
graph. Now, to solve \DS on a given graph~$\graph$ we can create the
graph~$\graph'=(A\cup B,E')$ as above, and solve \ST on $\graph'$ with $T=B$.
It is routine to check that any set $W$ that is a solution to \ST on $\graph'$
corresponds to a dominating set on $\graph$, and vice versa.
\medskip

\begin{figure}[t]
\begin{center}
\includegraphics[width=0.8\textwidth]{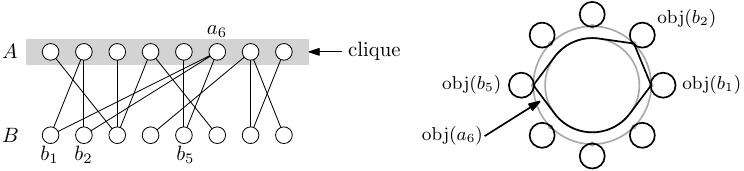}
\end{center}
\caption{Realizing a split graph as an intersection graph of fat objects.
The circles $\gamma_{\myin}$ and $\gamma_{\myout}$, and the points $p_i$,
are depicted in grey. For clarity, the objects $\mbox{obj}(a_i)$ for
$i\neq 6$ have been omitted.}\label{fig:fatpolygons}
\end{figure}

It remains to show that $\graph'$ is realizable as an intersection graph of
$(1-\eps)$-fat objects in $\Reals^2$. Let $\gamma_{\myout}$ and $\gamma_{\myin}$ be
two concentric circles, where $\gamma_{\myout}$ has radius~1 and $\gamma_{\myin}$
has radius~$(1-\eps)$. Let $\{p_1,\dots,p_n\}$ be $n$ points equally
spaced around $\gamma_{\myout}$. The object $\mbox{obj}(b_i)$ we create for
each vertex $b_i\in B$ is a disk of radius $1/n$ that touches $\gamma_{\myout}$
from the outside at $p_i$. The object $\mbox{obj}(a_i)$ we create for
each vertex $a_i\in A$ is the convex hull of the set
$\gamma_{\myin} \cup \{p_j\in P : (a_i,b_j) \in E'\}$;
see Fig.~\ref{fig:fatpolygons}.
Thus $\gamma_{\myin} \subset \mbox{obj}(a_i)\subset \gamma_{\myout}$,
and $\mbox{obj}(a_i)$ touches an object $\mbox{obj}(b_i)$ if and only
if $a_i$ is connected to $b_i$ in $\graph'$. Since the objects
$\mbox{obj}(b_i)$ are pairwise disjoint, and the objects $\mbox{obj}(a_i)$
all intersect, the intersection graph of the created objects equals~$\graph'$.
Furthermore, all objects are (at least) $(1-\eps)$-fat.
\end{myproof}

\paragraph{Remark.} The objects in the above construction are disks and convex objects
whose boundary consists of line segments and circular arcs. The construction can
also be done with objects that are convex polygons with $O(n+1/\eps)$ vertices.

\section{A lower bound for \DS in  ball graphs}
\label{se:LB-DS}
In this section we prove that under the ETH, there is no subexponential algorithm for \DS on ball graphs
in $\Reals^d$, for $d\geq 3$. It suffices to prove this for $d=3$, since any
ball graph in $\Reals^d$ can trivially be realized as a ball graph in $\Reals^{d+1}$.
We will use a reduction from a special version of \threeSAT, namely \niceSAT.
In a \niceSAT problem the input formula that we want to test for satisfiability is a
$(3,3)$-CNF formula, that is, a CNF formula in which every clause has at most
three literals, and every variable occurs at most three times in total.
\begin{proposition} {\rm\textbf{(De~Berg~\etal\cite{bbkmz-fetht-18})}}
There is no $2^{o(n)}$ algorithm for \niceSAT, unless the Exponential-Time Hypothesis fails.
\end{proposition}
Our reduction from \niceSAT to \DS on ball graphs works in two steps. First we
convert the given \niceSAT instance~$\phi$ to a graph $\graph_{\phi}$ that has a dominating
set of a certain size if and only if $\phi$ is satisfiable, and then we
realize $\graph_{\phi}$ as a ball graph in~$\Reals^d$.

\begin{figure}[t]
\centering
\includegraphics{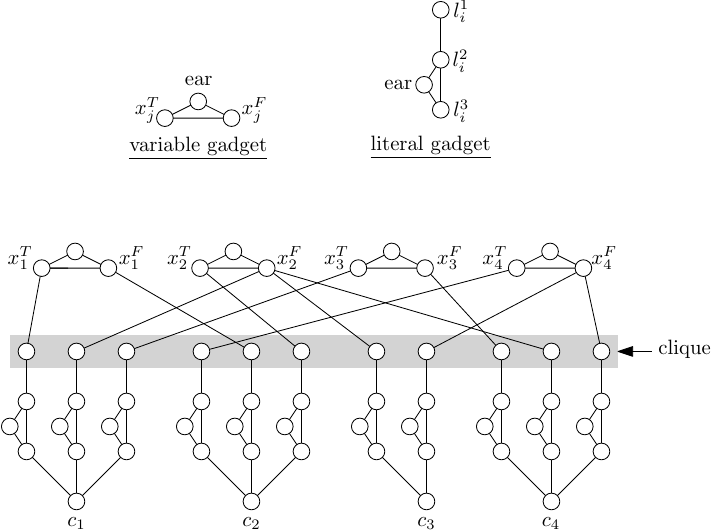}
\caption{Top: A variable gadget and a literal gadget.   Bottom: The graph $\graph_\phi$ for the formula $\phi := (x_1\vee \neg x_2\vee x_3)\wedge (x_4\vee \neg x_1\vee x_2) \wedge (\neg x_2 \vee \neg x_4) \wedge (\neg x_3\vee \neg x_2\vee x_4)$; edges in the clique on the vertices $l_i^1$ are omitted for clarity.}\label{fig:ball_graph}
\end{figure}

\paragraph{Step~1: Construction of~$\graph_{\phi}$.}
Let $\phi$ be a \niceSAT formula. We use a preprocessing step on $\phi$ in
order to remove clauses that have only one literal the following way. If a
clause has only one literal, we set its variable to satisfy the clause and
delete any newly satisfied clauses or false literals that were created. We
repeat this procedure as long as there are still clauses with only one
literal. If at some point all literals are deleted from a clause, then that
clause cannot be satisfied and we have solved the problem. In this case the
reduction algorithm would return a graph~$\graph_\phi$ and number~$k$ that
forms a trivial {\sc no}-instance. If all clauses are deleted because they are
satisfied, then we are done as well, and we return a trivial {\sc
yes}-instance. If none of these two cases arises then we are left with a
formula that has clauses of size two and three only, and that is satisfiable
if and only if the original formula was satisfiable. With a small abuse of
notation we still denote this formula by~$\phi$,

Let $\cX:=\{x_1,\ldots,x_n\}$ be the set of variables occurring in~$\phi$,
let $\cC := \{c_1,\ldots,c_s\}$ be the set of clauses in~$\phi$,
and let $\cL:=\{l_1,\ldots,l_t\}$ be the multiset of literals occurring in~$\phi$.
($\cL$ is a multiset because the same literal can occur multiple times
in~$\phi$, as for example $\neg x_2$ does in Fig.~\ref{fig:ball_graph}.)
The graph~$\graph_{\phi}$ is now constructed as follows;
see Fig.~\ref{fig:ball_graph} for an illustration.
\begin{itemize}
\item  For each variable $x_j\in \cX$ we create a \emph{variable gadget}, which consists
       of three vertices forming a triangle. Two of these vertices,
       labeled $x_j^T$ and $x_j^F$, correspond to
       truth assignments to~$x_j$: selecting $x_j^T$ into the dominating set
       corresponds to setting $x_j := \mytrue$, and selecting $x_j^F$ into the dominating
       set corresponds to setting $x_j := \myfalse$. The third vertex is called the
       \emph{ear} of the gadget.
\item For each clause $c_k\in \cC$ we create a \emph{clause gadget}, consisting
      of a single vertex labeled~$c_k$.
\item For each literal $l_i\in \cL$ we create a \emph{literal gadget}, which is a path of three
      vertices, labeled $l_i^1$, $l_i^2$, and $l_i^3$, plus an extra vertex---the \emph{ear} of the literal gadget---connected to $l_i^2$ and $l_i^3$.
\item The literal gadgets are connected to the variable and clause gadgets as follows.
      Let $c_k$ be the clause containing the literal~$l_i$ and let $x_j$ be the variable
      corresponding to $l_i$. Then we add the edge $(c_k,l_i^3)$, and we add $(x_j^T,l_i^1)$
      if $l_i = x_j$ and $(x_j^F,l_i^1)$ if $l_i=\neg x_j$.
\item Finally, we add edges between any pair of vertices in the set $\{l_i^1 : l_i\in\cL\}$,
      thus creating a clique on these vertices.
\end{itemize}
\begin{lemma}
$\graph_{\phi}$ has a dominating set of size $n+t$ if and only if $\phi$ is satisfiable.
\end{lemma}
\begin{myproof}
First suppose $\phi$ has a satisfying assignment. We can construct
a dominating set~$D$ of size $n+t$ for $\graph_{\phi}$ as follows.
For each variable $x_j\in \cX$ we add $x_j^T$ to $D$ if $x_j=\mytrue$, and we
add $x_j^F$ to $D$ if $x_j=\myfalse$. In addition, for each literal $l_i\in\cL$
we add $l_i^2$ to $D$ if $l_i=\myfalse$, and we add $l_i^3$ to $D$ if $l_i=\mytrue$.
The resulting set~$D$ has size $n+t$ since we have $n$ variables and $t$ literal
occurrences. To see that $D$ is a dominating set, observe that any vertex
in a variable gadget is dominated by the vertex selected from that gadget,
and observe that the vertices $l_i^2$, $l_i^3$ and the ear in the gadget for literal~$l_i$
are all dominated by the vertex selected from that gadget.
Vertex $l_i^1$ is also dominated, namely by a vertex in a variable
gadget (if $l_i=\mytrue$) or by $l_i^2$ (if $l_i=\myfalse$).
Finally, the vertex for any clause~$c_k$ must be dominated because
at least one literal $l_i$ in $c_k$ is true, meaning that $l_i^3\in D$.

Conversely, let $D$ be an arbitrary dominating set of size $n+t$.
The $n+t$ ears of $\graph_\phi$ have disjoint closed neighborhoods,
so to dominate all ears the set $D$ must contain exactly one vertex of each
of these closed neighborhoods.
Moreover, if $D$ contains an ear $v$, then we can exchange $v$ for one
of its neighbors $w$; the resulting set will still be dominating as $w$
dominates a superset of the vertices dominated by~$v$. Consequently, we
may assume that $D$ contains exactly one
vertex from $\{x_j^T,x_j^F\}$ for each $1\leq j\leq n$, and exactly one
vertex from $\{l_i^2,l_i^3\}$ for each $1\leq i\leq t$.
We now set $x_j:=\mytrue$ if and only if $x_j^T\in D$, for all $1\leq j\leq n$.
We claim that this assignment satisfies $\phi$. To see this, consider
a clause~$c_k$. Since the vertex labeled~$c_k$ must be dominated,
there is at least one literal~$l_i$ occurring in $c_k$ such that $l_i^3\in D$.
Suppose that $l_i=x_j$; the argument for $l_i=\neg x_j$ is similar.
Since $l_i^3\in D$ we have $l_i^2\not\in D$, and since the vertex~$l_i^1$
must be dominated this implies~$x_j^T\in D$. Hence, $x_j$ has been set
to $\mytrue$, thus satisfying~$c_k$.
\end{myproof}

\paragraph{Step~2: Realizing~$\graph_{\phi}$ as a ball graph.}
Next we show that the graph~$\graph_{\phi}$ obtained in Step~1 can be realized
as a ball graph in~$\Reals^3$. We need the following lemma.
\begin{lemma}\label{lem:skewlines}
Let $\ell_1$ be the $x$-axis
and let $\ell_2$ be the line $(0,0,h)+\lambda(0,1,0)$, for some $h\in\Reals$.
Let $p_1 := (x,0,0)$ and $p_2 := (0,y,h)$ be arbitrary points on $\ell_1$ and $\ell_2$,
respectively. Then there is a unique ball that touches
$\ell_1$ at $p_1$ and $\ell_2$ at~$p_2$. The center of this ball is $(x,y,z_{xy})$
and its radius is $r_{xy} \eqdef \sqrt{y^2+z_{xy}^2}$,
where $z_{xy}\eqdef \frac{h^2+x^2-y^2}{2h}$.
\end{lemma}
\begin{myproof}
Note that the center of a ball touching $\ell_1$ at $p_1$ and $\ell_2$ at $p_2$
must lie on each of the following planes:
\begin{enumerate}
	\item[$h_1$:] the plane through $p_1$ perpendicular to $\ell_1$
	\item[$h_2$:] the plane through $p_2$ perpendicular to $\ell_2$
	\item[$h_3$:] the perpendicular bisector plane of $p_1$ and $p_2$.
\end{enumerate}
Since $\ell_1$ and $\ell_2$ are skew lines---they do not intersect and are not
parallel---these three planes are pairwise non-parallel. Hence, they have a unique
intersection point $p$, and the ball with center $p$ and
radius $|pp_1|$ has the desired properties. Verifying that the coordinates of
the center and the radius
of this ball are as claimed, is a routine computation which we omit.
\end{myproof}
Next we show how to realize the graph~$\graph_\phi$ as a ball graph in~$\Reals^3$.
Without loss of generality, assume that the literal occurrences and clauses in the formula~$\phi$
are indexed left to right, so that for example $c_1$ is the first clause and $l_1$ is the first
literal in $c_1$, and $c_k$ is the last clause and $l_t$ is the last literal in~$c_k$.

\begin{figure}[t]
\centering
\includegraphics[width=\textwidth]{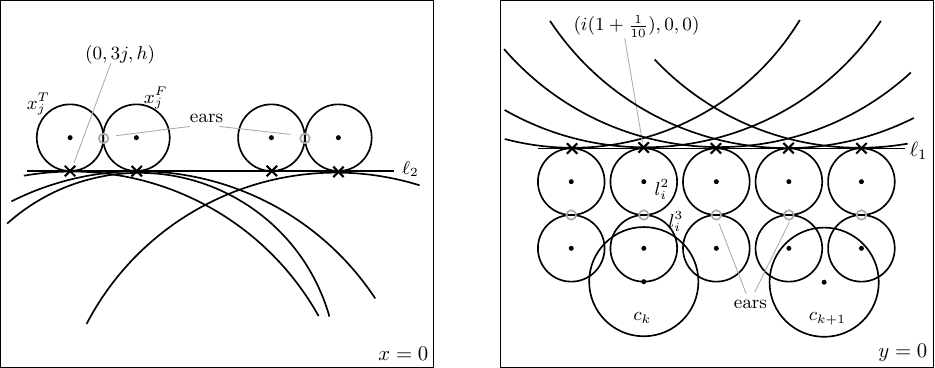}
\caption{Cross-section of the construction in the plane $x=0$ (left) and $y=0$ (right). Note that the large balls $l_i^1$ have their centers outside these planes. On the left, we have a literal $\neg x_j$ that occurs twice.}\label{fig:cross_sec}
\end{figure}

Let the line $\ell_1$ be the $x$-axis, and let $\ell_2$ be the line $(0,0,h)+\lambda(0,1,0)$,
where the height $h$ will be defined later. The idea is that the balls for the vertices
$x_j^T$ and $x_j^F$ will be touching $\ell_2$ from above, the balls for the vertices ~$l_i^2$
will be touching $\ell_1$ from below, and that Lemma~\ref{lem:skewlines} will then allow us to
place the balls for $l_i^1$ such that the correct connections are realized.
Next we describe in detail how the balls representing the vertices of $\graph_{\phi}$
are placed. We denote the ball representing a node~$v$ by $\ball(v)$, so for instance
$\ball(c_k)$ denotes the ball representing the vertex in the gadget for clause~$c_k$.
\begin{itemize}
\item The vertices labeled $x_j^T$ and $x_j^F$ of the variable gadget for~$x_j$ will
      be represented by balls $\ball(x_j^T)$ and $\ball(x_j^F)$ with centers $(0,3j,h+\frac{1}{2})$ and $(0,3j+1,h+\frac{1}{2})$, respectively.
      These balls have radius~$\frac{1}{2}$, so that they touch~$\ell_2$ at the points $(0,3j,h)$ and $(0,3j+1,h)$. Note that $\ball(x_j^T)$ and $\ball(x_j^F)$ touch each other at the point~$(0,3j+\frac{1}{2},h+\frac{1}{2})$.
      The ear of the gadget for $x_j$ is a ball with that touching point as center, and with a sufficiently small radius, say~$\frac{1}{20}$; see Fig.~\ref{fig:cross_sec}(left).
      Observe that the intersection graph of these three balls is a triangle, as required.
\item For a clause~$c_k$, the center of $\ball(c_k)$ is
      $(i^*(1+\frac{1}{10}),0,-2)$ and its radius is~9/10,  where value of $i^*$
      depends on whether $c_k$ has two or three literals. In the former case,
      $i^* = i+\frac{1}{2}$, where $i$ is the index such that $c_k$ contains literals~$l_i$ and~$l_{i+1}$,
      in the latter case~$i^*=i$, where $i$ is the index such that $c_k$ contains
      literals~$l_{i-1}$, $l_i$, and~$l_{i+1}$.
\item The balls representing the vertices in the gadget for literal~$l_i$ are defined as follows.
      The balls $\ball(l_i^2)$ and $\ball(l_i^3)$ have radius $\frac{1}{2}$ and centers $(i(1+\frac{1}{10}),0,-\frac{1}{2})$ and $(i(1+\frac{1}{10}),0,-\frac{3}{2})$, respectively.
      Note that if $c_k$ is the clause containing literal~$l_i$, then $\ball(l_i^3)$
      intersects $\ball(c_k)$; see Fig.~\ref{fig:cross_sec}(right).
      Also note that the balls $\ball(l_i^2)$ and $\ball(l_i^3)$ touch at the point $(i(1+\frac{1}{10}),0,-1)$. The ball representing
      the ear of the gadget will be centered at that touching point and have
      radius~$\frac{1}{20}$. Thus the
      intersection graph of $\ball(l_i^2)$, $\ball(l_i^3)$ and the ball representing the ear
      is a triangle, as required.
      \\[2mm]
      The crucial part of the construction is the placement of $\ball(l_i^1)$, which
      is done as follows. We require~$\ball(l_i^1)$ to touch~$\ell_1$ at the
      point~$(i(1+\frac{1}{10}),0,0)$. Note that $\ball(l_i^2)$ also touches~$\ell_1$
      at this point, thus we have realized the edge~$(l_i^1,l_i^2)$ in~$\graph_{\phi}$.
      We also require $\ball(l_i^2)$ to touch the line~$\ell_2$. The touching point depends
      on whether $l_i$ is a  positive or a negative literal:
      if $l_i=x_j$ then we require $\ball(l_i^1)$ to touch $\ell_2$ at $(0,3j,h)$,
      and if $l_i=\neg x_j$ then we require $\ball(l_i^1)$ to touch $\ell_2$ at $(0,3j+1,h)$.
      In the former case it will intersect $\ball(x_j^T)$ and in the latter case it will
      intersect~$\ball(x_j^F)$, as required. By Lemma~\ref{lem:skewlines} there is a unique
      ball touching $\ell_1$ and $\ell_2$ as just specified.
\end{itemize}
Let $S_\phi$ be the collection of balls just defined. The next lemma states that our construction
realizes the graph~$\graph_\phi$ if we pick the value~$h$ specifying the $x$-coordinate of~$\ell_2$
appropriately.
\begin{lemma}
Let $h:=20N^2$, where $N :=\max(3n+1,(1+\frac{1}{10})t)$.
Then the intersection graph $\graph[S_\phi]$ induced by~$S_\phi$ is
the graph~$\graph_\phi$.
\end{lemma}
\begin{myproof}
We already argued above that most edges of $\graph_{\phi}$ are also present
in $\graph[S_\phi]$, that is, if $(u,v)$ is an edge in $\graph_\phi$ then
$\ball(u)$ intersects~$\ball(v)$. The only exception are the edges forming
the clique on the set $\{l_i^1: 1\leq i \leq t\}$. Thus it remains to show
that these edges are present as well,\footnote{Actually, it turns out that one can also argue that the existence of
these edges is not needed for the reduction to work.
We prefer to work with the specific graph~$\graph_\phi$
defined earlier, and therefore need to show that our
geometric representation includes all edges in the clique.}
and that no spurious edges are present in~$\graph[S_\phi]$. It is straightforward
to verify that any spurious edge that might arise must involve a ball
$\ball(l_i^1)$.

Consider a ball~$\ball(l_i^1)$. Recall that $\ball(l_i^1)$ touches~$\ell_1$ at the point~$(i(1+\frac{1}{10}),0,0)$,
and $\ell_1$ at $(0,3j,h)$ or $(0,3j+1,h)$ for some $1\leq j\leq n$ (depending
on whether $l_i$ is a positive or negative literal).
Note that the first two coordinates of these touching points are in the interval $[0,N]$,
and that we set $h := 20 N^2$.
Recall from Lemma~\ref{lem:skewlines} that the ball touching $\ell_1$ at $(x,0,0)$ and $\ell_2$ at $(0,y,h)$ has center $(x,y,z_{xy})$ and radius is $r_{xy} \eqdef \sqrt{y^2+z_{xy}^2}$,
where $z_{xy}\eqdef \frac{h^2+x^2-y^2}{2h}$. Thus for the $z$-coordinate of the center of~$\ball(l_i^1)$
we have
\[
z_{xy}=\frac{h^2+	x^2 - y^2}{2h} \in \left[ \frac{h^2-N^2}{2h},\frac{h^2+N^2}{2h}\right]= \left[\frac{h}{2}-\frac{1}{40}, \frac{h}{2}+\frac{1}{40}\right],
\]
and for the radius of $\ball(l_i^1)$ we have
\[
\begin{array}{lll}
\radius(\ball(l_i^1)) & = & \sqrt{y^2+z_{xy}^2}\leq\sqrt{N^2+\left(\frac{h}{2}+\frac{1}{40}\right)^2} \\[2mm]
  & = & \sqrt{\frac{h}{20}+\left(\frac{h}{2}\right)^2+\frac{h}{40}+\frac{1}{1600}} \\[2mm]
  & < & \frac{h}{2} + \frac{1}{10}.
\end{array}
\]
Consequently, the ball is disjoint from both of the open halfspaces $z>h+1/5$ and $z<-1/5$.
These open half-spaces contain all the balls corresponding to ears and all balls $\ball(c_k)$
and $\ball(l_i^3)$. Now consider two balls $\ball(l_i^1)$ and $\ball(l_{i'}^1)$.
The distance between the centers of these balls, $(x,y,z_{xy})$ and $(x',y',z_{x'y'})$,
 is at most
 \[
 (x-x')^2+(y-y')^2+(z_{xy}-z_{x'y'})^2\leq 2N^2+\left(\frac{1}{20}\right)^2< h/3 < r_{xy}+r_{x'y'}.
 \]
 Therefore, the set $\{\ball(l_i^1) :  1\leq i\leq t\}$ induces a clique, as desired.

It remains to be shown that $\ball(l_i^1)$ is disjoint from any ball
$\ball(l_{i'}^2)$ for $i'\neq i$, and from any ball $\ball(x_j^T)$ and $\ball(x_j^F)$
except for the ball touching $\ell_2$ at the same point as $\ball(l_i^1)$ touches~$\ell_2$.
Let $(x,y,z_{xy})$ be the center of $\ball(l_i^1)$ and
consider the distance of $(x,y,z_{xy})$ to a point $(x',0,-\frac{1}{2})$, where $|x-x'|\geq 1$.
(The argument is analogous for the distance of $(x,y,z_{xy})$ to $(0,y',h+\frac{1}{2})$, where $|y-y'|\geq 1$.) It is sufficient to show that this distance is larger than the sum of the ball radii, that is, we need that $\dist((x,y,z_{xy}),(x',0,-\frac{1}{2})) > r_{xy}+\frac{1}{2}$. Taking the squared distance instead, we need to show that
\[
(x'-x)^2+y^2+\left(z_{xy}+\frac{1}{2}\right)^2 > \left(\sqrt{y^2+z_{xy}^2} + \frac{1}{2} \right)^2,
\]
which is equivalent to
\[
(x'-x)^2 + z_{xy} > \sqrt{y^2+z_{xy}^2}.
\]
Since $y\leq N$, we have that $y^2\leq N^2=\frac{h}{20} < \frac{z_{xy}}{8}$, so we can bound the left hand side as:
\begin{align*}
\sqrt{y^2+z_{xy}^2} < \sqrt{\frac{z_{xy}}{8} + z_{xy}^2}<z_{xy}+\frac{1}{4}< z_{xy} + (x'-x)^2,
\end{align*}
where the last inequality follows since $x$ and $x'$ are distinct integers, therefore each $\ball(l_i^1)$ has the correct intersections.

Finally, although the above construction  uses some irrational ball radii, it is sufficient to use rational radii of precision $O(1/n)$, which results in the same intersection graph.
\end{myproof}

Putting it all together we obtain the main result of this section.
\begin{theorem}\label{thm:ballgraph}
There is no $2^{o(n)}$ algorithm for \DS in ball graphs, unless the Exponential-Time Hypothesis fails.
\end{theorem}

\section{A lower bound for \wDS in  unit-ball graphs}
\label{se:LB-wDS}
We now turn our attention to \wDS for ball graphs in~$\Reals^3$.
Here we can even prove lower bounds for unit-ball graphs.
\medskip

Our reduction is again from $(3,3)$-SAT.
Let $\phi$ be a $(3,3)$-SAT formula, which we preprocess so that all clauses have size two or three as in the beginning of the proof of Theorem~\ref{thm:ballgraph}. First, we will create a vertex-weighted graph $\graph_{\phi}$ that has a dominating set of a given size $k$ if and only if $\phi$ is satisfiable.
 Then, we show that this graph can be realized as the intersection graph of unit balls in $\Reals^3$.

\begin{figure}[t]
\centering
\includegraphics{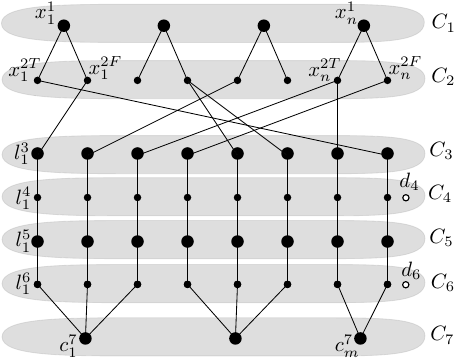}
\caption{The weighted graph $\graph_{\phi}$. Vertices in each group $C_i$ form a clique. Large nodes have infinite weight, the dummies $d_4$ and $d_6$ have weight $0$, and all other nodes have weight $1$.}\label{fig:weighted_graph}
\end{figure}

\paragraph{Step~1: Construction of~$\graph_{\phi}$.}
Let $\phi$ have variables $x_1,\dots,x_n$, literals $l_1,\dots,l_t$ (as a multiset), and clauses $c_1,\dots,c_m$. We create a vertex-weighted graph $\graph_{\phi}$ that consists of seven cliques, denoted by $C^1,\dots,C^7$. The remaining (non-clique) edges of the graph will all be between two consecutive cliques $C^p$ and $C^{p+1}$. Clique $C^1$ has $n$ vertices labeled $x_j^1$, one corresponding to each variable. Clique $C^2$ has $2n$ vertices, where for each variable we have two vertices corresponding to setting the variable to true or false. These are labeled $x_j^{2T}$ and $x_j^{2F}$, see Figure~\ref{fig:weighted_graph}. Cliques $C^3,\dots,C^6$ each contain a single vertex for each literal occurrence $l_i$, labeled $l_i^3,l_i^4,l_i^5,l_i^6$. Finally, $C^7$ has a single vertex for each clause $c_k\, (k=1,\dots,m)$, labeled as $c_k^7$. In addition to these, we add a dummy vertex to $C^4$ and $C^6$, labeled $d^4$ and $d^6$ respectively. Apart from the edges in the cliques, the other edges are defined as follows. For $j=1\dots,n$, we add the edges $x_j^1x_j^{2T}$ and $x_j^1x_j^{2F}$. We connect the first literal vertices to the corresponding variable setting, i.e., for each positive literal $l_i=x_j$, we add $x_j^{2T}l_i^3$, and for each negative literal $l_i=\neg x_j$, we add $x_j^{2F}l_i^3$. For each literal $l_i$ we add the edges of the path $l_i^3l_i^4l_i^5l_i^6$. Finally, for $k=1,\dots,m$, we connect $c_k^7$ to $l_i^6$ if and only if $l_i$ is occurs in $c_k$. The vertex weights in $C^1,C^3,C^5$, and $C^7$ are set to $\infty$, the dummy vertices get weight\footnote{Alternatively, one could give the dummies a weight of $1$, and add a unique neighbor to each of them with weight $\infty$, which ensures that they are contained in all finite weight dominating sets.} $0$, and all other vertex weights are set to $1$.

\begin{lemma}
The graph $\graph_\phi$ has a dominating set of weight $n+t$ if and only if $\phi$ is satisfiable.
\end{lemma}

\begin{myproof}
First suppose that $\phi$ is satisfiable. We show how to
create a dominating set $D$ of weight $n+t$ from a satisfying assignment. For each true variable $x_j$, we add the vertex $x_j^{2T}$ to $D$; for each false variable $x_j$, we add $x_j^{2F}$ instead. For true literals, we add $l_i^6$ to $D$; for false literals we add $l_i^4$  to $D$. Finally, we add the two dummies.
Note that the weight of $D$ is $n+t$.
To see that every vertex is dominated, first observe that vertices in $C^1$ are dominated since exactly one of their neighbors is in $D\cap C^2$, and vertices in $C^7$ are dominated since at least one of their literals is true, i.e., the corresponding vertex in $C^6$ is in $D$. Furthermore, there is at least one vertex selected in $C^2$, so $C^2$ is dominated as well. The dummies ensure that $C^4$ and $C^6$ are dominated. Finally, all other vertices are on some path $x_j^{2.}l_i^3l_i^4l_i^5l_i^6$, where either $x_j^{2.}$ and $l_i^6$ are selected (for true literals) or $l_i^4$ is selected (for false literals), which means that $C^3$ and $C^5$ are dominated as well.
\medskip

Now suppose $\graph_{\phi}$ has a dominating set $D$ of weight $n+t$. Based on $D$, we show how to construct a satisfying assignment of~$\phi$. Note that $D$ cannot contain vertices of infinite weight. Consequently, all vertices in $C^1,C^3,C^5$, and $C^7$ must be dominated from outside. Note that the closed neighborhood of each vertex in $C^1\cup C^5$ consists of their own clique and two other vertices that we can call potential dominators: indeed, $x_j^1$ can only be dominated by $x_j^{2T}$ or $x_j^{2F}$, and $l_i^5$ can only be dominated by $l_i^4$ or $l_i^6$. Since there are $n+t$ vertices in $C^1 \cup C^5$, and all their potential dominators are disjoint sets, it follows that $D$ consists of exactly one vertex from each pair $\{x_j^{2T},x_j^{2F}\}\, j=1\dots n$ and $\{l_i^4,l_i^6\}\, i=1\dots t$. Consider the assignment where $x_i$ is set to true if and only if $x_j^{2T} \in D$. Suppose for the purpose of contradiction that a clause $c_k$ is not satisfied. Let $i$ be the index of a literal occurrence for which $l_i^6\in D$; such an index exists since $c_k^7$ can only be dominated by a vertex from $C^6$. Consequently, $l_i^4 \not\in D$, so in order to dominate $l_i^3$, it must be the case that its potential dominator in $C^2$ is in $D$. That is, if $l_i=x_j$, then $x_j^{2T}\in D$, and if $l_i=\neg x_j$, then $x_j^{2F}\in D$. But this would mean that $c_k$ is satisfied by this literal in our assignment, which is a contradiction.
\end{myproof}

\begin{figure}[t]
\centering
\includegraphics[height=6cm]{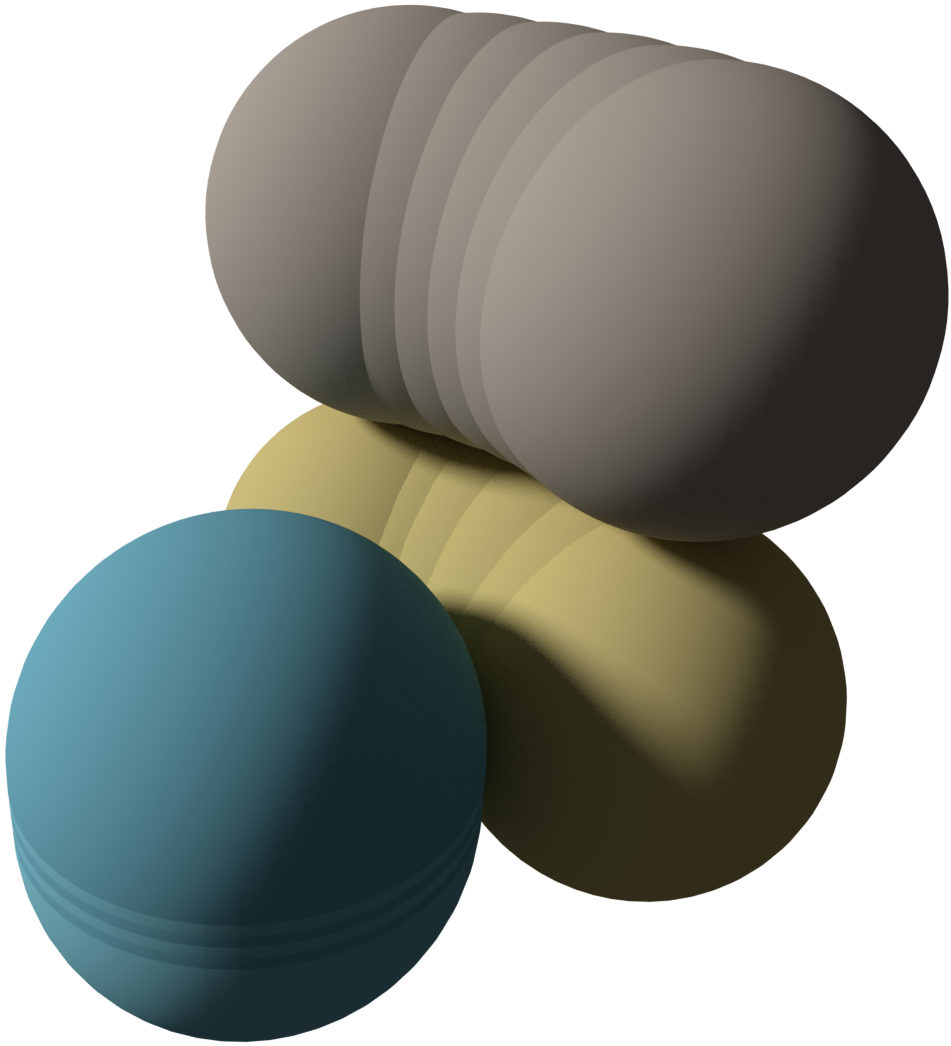}%
\hfill%
\includegraphics[height=6cm]{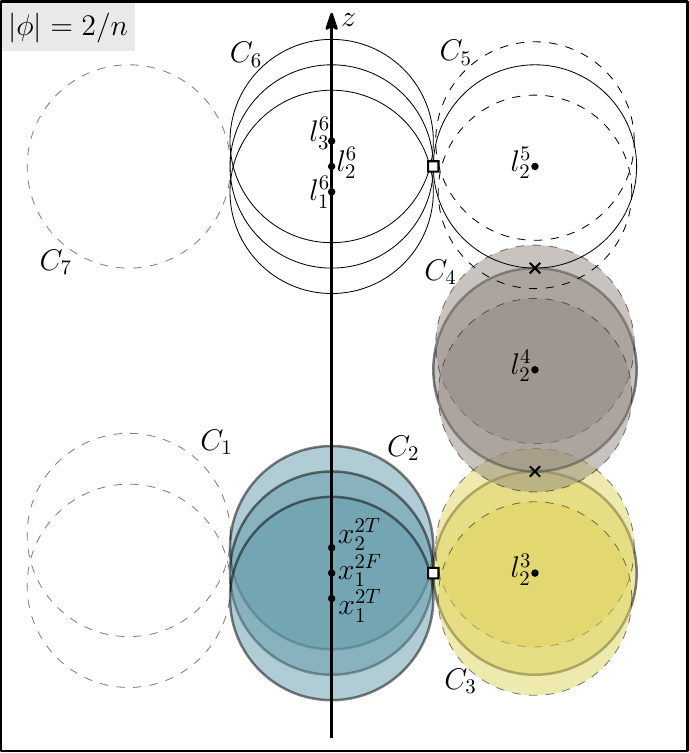}
\caption{Left: cliques $C_2,C_3$ and $C_4$ with blue, yellow and gray colors. The ball centers of $C_2$ are very close to each other so alternating shades of blue are used to distinguish the balls. Right: the intersection of the first few balls from each clique with the plane $|\phi|=2/n$. Balls from different cliques have two touching points (indicated by squares) and two non-empty intersections (indicated by crosses) within this plane. The smaller dashed disks correspond to balls whose center is outside this plane.}\label{fig:weightedrender}
\end{figure}

As in the proof of Theorem~\ref{thm:ballgraph}, the size of $V(\graph_\phi)$ is $O(n)$, so it suffices to show that $\graph_\phi$ can be realized as a unit ball graph.
(For simplicity we used infinite weights in the construction, but we can use weights that
are polynomial in the size of the instance as well.)

\paragraph{Step~2: Realizing~$\graph_{\phi}$ as a unit ball graph.}

\begin{lemma}
The graph $\graph_\phi$ can be realized as a unit ball graph.
\end{lemma}

\begin{myproof}
It will be convenient to use cylindrical coordinates: the point $(x,y,z)$
becomes $(r,\phi,z)$, where $r=\sqrt{x^2+y^2}$ and $\phi=\mathrm{arctan}(y/x)$.
We will first define the ball centers for cliques $C^2$ through $C^6$,
and then for the cliques~$C^1$ and $C^7$. The balls in our construction will all have
diameter~1 (rather than radius~1), so that two balls intersect if and only if their
distance is at most~$1$.

Let $\eps>0$ be a small number that will be determined later. We place the centers
of the balls for $C^2$ in the natural order close to each other on the $z$  axis.
More precisely, we assign the (cylindrical) coordinates $(0,0,(2j-1)\eps)$ to the center of $\ball(x_j^{2T})$
and $(0,0,2j\eps)$ to  the center of $\ball(x_j^{2F})$. Without loss of generality,
assume that the literal occurrences and clauses are indexed left to right, so that
$l_1$ is the first literal in $c_1$, and $l_t$ is the last literal in the last clause $c_k$.
We now define the centers of the balls for $C^3$. If $l_i=x_j$, then the center of $\ball(l_i^3)$
is $\cent(l_i^3):=(1,i/n,(2j-1)\eps)$; if $l_i=\neg x_j$, then we set $\cent(l_i^3):=(1,i/n,2j\eps)$.
Observe that all balls of $C^3$ touch the cylinder $r=\frac{1}{2}$ at a single point,
which is also a touching point with the correct ball from $C^2$, see Figure~\ref{fig:weightedrender}.
(Note that the picture overestimates the distances within $C_2$; with the correct placement the balls are indistinguishable to the naked eye.)

Next, we set $\cent(l_i^4):=(1,i/n,1)$ and $\cent(l_i^5):=(1,i/n,2-i\eps)$. By choosing $\eps$ small enough, we can ensure that each ball $\ball(l_i^4)$ intersects only $\ball(l_i^3)$ from $C^3$ and $\ball(l_i^5)$ from $C^5$. To see this, note that the squared distance of a center $(1,\phi,z)$ to a center $(1,\phi',z')$ is
\[
(1-\cos(\phi-\phi'))^2+\sin^2(\phi-\phi')+(z-z')=2-2\cos(\phi-\phi')+(z-z')^2.
\]
If $\phi=\phi'$ and $z-z'<1$, then this distance is less than~1, so that $\ball(l_i^4)$ does intersect $\ball(l_i^3)$ and $\ball(l_i^5)$. Otherwise, the distance of $\cent(l_i^4)$ to any of the centers in $C^3\cup C^5$ is minimized if $\phi-\phi'=1/n$ and $z-z'=1-t\eps$. To avoid unwanted intersections, we need the following:
\[
 2-2\cos(1/n)+(1-t\eps)^2 > 1
\]
which is equivalent to
\[
 1-t\eps+\frac{t^2\eps^2}{2} > \cos(1/n).
\]
We set $\eps=1/(3tn^2)$. Note that this implies $\eps= \Theta(1/n^3)$. By the power series of cosine, we have
\[\cos(1/n)<1-\frac{1}{2n^2}+\frac{1}{24n^4}<1-\frac{1}{3n^2}=1-t\eps<1-t\eps+\frac{t^2\eps^2}{2},\]
so $\ball(l_i^4)$ does not intersect $\ball(l_j^3)$ and $\ball(l_j^5)$ for any $i\neq j$.

We continue with placing the rest of the centers. Let $\cent(l_i^6):=(0,0,2-i\eps)$ so that $\ball(l_i^6)$ touches $\ball(l_i^5)$; a similar structure was used between $C^3$ and $C^2$. We set $\cent(x_j^1):=(r,\pi,(2j-\frac{3}{2})\eps)$, where $r=\sqrt{1-\eps^2/4}$ is chosen so that $\ball(x_j^1)$ touches $\ball(x_j^{2T})$ and $\ball(x_j^{2F})$. Moving on to the placements for $C_7$, notice that literal occurrences of any given clause correspond to consecutive ball centers in $C_6$. We set $\cent(c_k^7):=(r,\pi,z)$, where $r$ and $z$ are defined the following way. The height is $z=2-i\eps$ if $l_i$ is the middle literal of $c_k$, and it is $z=-(i+\frac{1}{2})\eps$ if $c_k=(l_i \vee l_{i+1})$. The radius $r$ is chosen to ensure that the ball of $c_k$ touches the ball of its first and last literal: we set $r=\sqrt{1-\eps^2/4}$ and $r=\sqrt{1-\eps^2}$ for clauses of size two and three respectively. To conclude the construction, we place balls to represent the dummy vertices $d_4$ and $d_6$ at $\cent(d_4):=(1.5,0.5,1)$ and $\cent(d_6):=(0,0,2.5)$. It is routine to check that
this correctly realizes~$\graph_{\phi}$.

Finally, although the above construction  uses some irrational coordinates, one can use rational coordinates of precision $O(1/n^4)$.
\end{myproof}

The above construction and the lemmas imply the following theorem.

\begin{theorem}\label{thm:weighted}
There is no $2^{o(n)}$ algorithm for \wDS in unit ball graphs, unless the Exponential-Time Hypothesis fails.
\end{theorem}

\section{Concluding remarks}
De~Berg~\etal~\cite{bbkmz-fetht-18} recently presented a framework to solve
many classic graph problems in subexponential time on intersections graphs of
similarly-sized fat objects. We have shown that extending the framework in its
full generality to arbitrary (non-similarly-sized) fat objects is impossible.
More precisely, we have shown that \DS, one of the problems that the framework
for similarly-sized fat objects can handle, does not admit a subexponential
algorithm on arbitrary ball graphs in~$\Reals^3$, assuming ETH. Thus it seems
that obtaining subexponential algorithms for arbitrary fat objects, if
possible, will require rather problem-specific arguments. Similarly, obtaining
subexponential algorithms for weighted problems on similarly-sized objects is
not always possible: \wDS does not admit a subexponential algorithm on
unit-ball graphs in $\Reals^3$, assuming ETH. For arbitrary fat objects
(instead of just balls), the situation is worse: even in $\Reals^2$ the
problems \DS, \CDS, and \ST do not admit subexponential algorithms, assuming
ETH. On the positive side, we argued that if the solution to the problem at
hand (or its complement) can contain at most a constant number of vertices
from a clique, then the technique from De~Berg~\etal~\cite{bbkmz-fetht-18} may
be applicable; this is for example the case for \CVC, \FVS, and \CFVS.

Several questions are left open by our study. First and foremost, although we have ruled out a
subexponential algorithm for \DS in $\Reals^2$ for arbitrary fat objects, the complexity of \DS
on disk graphs remains open. Here, related work on \textsc{Geometric Set Cover}~\cite{facility,orthant}
leads us to believe that subexponential algorithms may be attainable for disks, possibly
even for the weighted version.
\begin{conjecture}
There is a $2^{\Otilde(\sqrt{n})}$ algorithm for \textsc{(Weighted)} \DS in disk graphs.
\end{conjecture}
A prominent problem studied in~\cite{bbkmz-fetht-18} that we did not tackle here is \HC.
Is there a subexponential algorithm for \HC in disk graphs? What about ball graphs?
Finally, the weighted versions of almost all of these problems remain open in $\Reals^2$,
as long as we have similarly sized objects. For example, what is the complexity of vertex weighted \ST
in unit-disk graphs?

\bibliographystyle{plain}
\newcommand{\cgta}{\emph{Comput.\ Geom.\ Theory Appl.}\xspace}
\newcommand{\dcg}{\emph{Discr.\ Comput.\ Geom.}\xspace}
\newcommand{\dm}{\emph{Discr.\ Math.}\xspace}
\newcommand{\ijcga}{\emph{Int.\ J.\ Comput.\ Geom.\ Appl.}\xspace}
\newcommand{\algor}{\emph{Algorithmica}\xspace}
\newcommand{\sicomp}{\emph{SIAM J.\ Comput.}\xspace}
\newcommand{\siap}{\emph{SIAM J.\ App.\ Math.}\xspace}
\newcommand{\jalg}{\emph{J.\ Alg.}\xspace}
\newcommand{\jda}{\emph{J.\ Discr.\ Alg.}\xspace}
\newcommand{\ipl}{\emph{Inf.\ Proc.\ Lett.}\xspace}
\newcommand{\jacm}{\emph{J.~ACM}\xspace}
\newcommand{\tcs}{\emph{Theoret.\ Comput.\ Sci.}\xspace}
\newcommand{\jcss}{\emph{J.\ Comput.\ Sys.\ Sci.}\xspace}

\newcommand{\socg}[1]{In \emph{Proc.\ #1 ACM Symp.\ Comput.\ Geom. (SoCG)}\xspace}
\newcommand{\soda}[1]{In \emph{Proc.\ #1 ACM-SIAM Symp.\ Discr.\ Alg. (SODA)}\xspace}
\newcommand{\stoc}[1]{In \emph{Proc.\ #1 ACM Symp.\ Theory Comp. (STOC)}\xspace}
\newcommand{\focs}[1]{In \emph{Proc.\ #1 IEEE Symp. Foundations Comp. Sci. (FOCS)}\xspace}
\newcommand{\esa}[1]{In \emph{Proc.\ #1 Europ.\ Symp.\ Alg. (ESA)}\xspace}
\newcommand{\swat}[1]{In \emph{Proc.\ #1 Scandinavian Workshop.\ Alg.\ Theory (SWAT)}\xspace}
\newcommand{\wg}[1]{In \emph{Proc.\ #1 Workshop.\ Graph Theoret. Concepts Comput. Sci. (WG)}\xspace}
\newcommand{\isaac}[1]{In \emph{Proc.\ #1 Int. Sympos. Alg. Comput. (ISAAC)}\xspace}
\newcommand{\icalp}[1]{In \emph{Proc.\ #1 Int. Coll. Automata, Languages, and Programming (ICALP)}\xspace}

\end{document}